\documentclass{article}

\usepackage[preprint, nonatbib]{nips_2018}

\usepackage{float}

\usepackage[square,numbers]{natbib}
\usepackage[utf8]{inputenc} % allow utf-8 input
\usepackage[T1]{fontenc}    % use 8-bit T1 fonts
\usepackage{hyperref}       % hyperlinks
\usepackage{url}            % simple URL typesetting
\usepackage{booktabs}       % professional-quality tables
\usepackage{amsfonts}       % blackboard math symbols
\usepackage{nicefrac}       % compact symbols for 1/2, etc.
\usepackage{microtype}      % microtypography

\usepackage{graphicx}
\usepackage{tabularx}
\usepackage{hyperref}
\usepackage{algorithm}
\usepackage{algorithmic}
\usepackage{framed,multirow}
\usepackage{amssymb}
\usepackage{latexsym}

\usepackage{multirow}
\usepackage{booktabs}
\usepackage{color}
\usepackage{natbib}
\usepackage{verbatim}
\usepackage{graphicx}
\usepackage{url}
\usepackage{xcolor}

\usepackage{geometry}

\title{Automatic Generation of Large-scale 3D Road Networks based on GIS Data}

\author{
  Hua Wang \\
  
\And
  Yue Wu \\
 
\And
  Xu Han \\
  
\And
  Mingliang Xu* \\
  
\And
  Weizhe Chen \\
  
\And
  Guoliang Chen \\
 
}

\begin{document}
\maketitle

\begin{abstract}
%%%
How to automatically generate a realistic large-scale 3D road network is a key point for immersive and credible traffic simulations. Existing methods cannot automatically generate various kinds of intersections in 3D space based on GIS data. In this paper, we propose a method to generate complex and large-scale 3D road networks automatically with the open source GIS data, including satellite imagery, elevation data and two-dimensional(2D) road center axis data, as input. We first introduce a semantic structure of road network to obtain high-detailed and well-formed networks in a 3D scene. We then generate 2D shapes and topological data of the road network according to the semantic structure and 2D road center axis data. At last, we segment the elevation data and generate the surface of the 3D road network according to the 2D semantic data and satellite imagery data. Results show that our method does well in the generation of various types of intersections and the high-detailed features of roads. The traffic semantic structure, which must be provided in traffic simulation, can also be generated automatically according to our method.

%%%%
\end{abstract}

\section{Introduction}
\label{sec1}
Three-dimensional(3D) virtual road network can give a high immersive visualization. It is important for traffic simulations and intelligent traffic controls \cite{Qianwen2020,HuaWang2020,allinOne2014,huawang2017,Yeh2015}.  The following two parts are crucial in a large-scale 3D virtual road network generation: 1) a realistic 3D scene with smooth road surface; 2) a traffic semantic structure of road network. How to automatically generate a realistic large-scale 3D road network including realistic scene, smooth road surfaces and traffic semantic structures based on the existing Geographic Information Systems (GIS) data (such as satellite imagery, elevation data, road network vector data) is a key point for immersive and credible traffic simulations \cite{CAO2020,WU2021,TSIOTAS2017}.

There are mainly two methods focusing on generating realistic 3D roads. One generates scenes by combining 3D models (road data, environment data and so on) into a unified view \cite{Garcia2017Fast,Parish2001,Dias20063D}. In order to generate a realistic scene, the data of those 3D models can be collected by some data acquisition devices, such as airborne tilt photography equipment\cite{KU202013}. However, there are still lots of difficulties for data denoising and texture reconstruction. The other method combines common GIS data to generate 3D roads \cite{Galin2010Procedural,Zhiguang20183D}. The satellite imagery is used as texture of roads. In this way, the generated scenes are based on the real data, so they are realistic enough. Nevertheless, it is difficult to automatically generate various kinds of intersections in 3D space. In addition, both of two methods cannot generate the road structure that can be used for traffic simulation well directly.

To tackle the aforementioned challenges, in this paper, we propose an automatic method to generate large-scale 3D virtual road networks based on GIS data. We first introduce a semantic description of the large-scale road networks, and then generate the 2D shape of the road network automatically according to the semantic description and geometric design standards of roads. At last, we segment the elevation data based on the 2D shape and use piecewise B-spline curves with monotone curvature in the Frenet frame to fit the elevation curves of roads. With the open source GIS data, including satellite imagery, elevation data and 2D road center axis data as input, our method can automatically generate a large-scale 3D virtual road network with realistic environment and high-detailed road surfaces. Traffic semantic structures such as connectivity and adjacency relationship and so on can also be generated automatically according to our method.

To sum up, the contributions of our work are summarized as follows:
\begin{itemize}
\vspace{-0.25cm}
\item a method of creating 3D scenes automatically with only open source GIS data(satellite imagery, elevation data and 2D road center axis data) as input. It is worth mentioning that our method supports for generating diverse and complex 3D intersections automatically.
\vspace{-0.25cm}
\item a way for segmenting and smoothing elevation data, so that the 3D road network has high-detailed road surface and it is also in accordance with the code of civil engineering.
\vspace{-0.25cm} 
\item a semantic structure for 3D road network that can be used in traffic simulation. 
\end{itemize}
For the rest of the paper, we ﬁrst review the literatures of 3D scene modeling and traffic network modeling in Section 2, and in Section 3 we introduce our method in detail. In Section 4, we propose a traffic semantic structure for traffic simulation. At last, we discuss the method and draw some conclusions in Section 5.

\section{Related work}
Our work relies on two parts: One is 3D scene modeling and the other is traffic network modeling.

\subsection{3D scene modeling}
Using 3D models to generate 3D scenes is a traditional method. Garciadorado et al. \cite{Garcia2017Fast} created roads automatically in 3D scenes by taking the attributes of lane number, intersections and the shape of roads as the input of the program. Parish et al. \cite{Parish2001} used the software named CityEngine to create expressways and lanes in a large-scale 3D urban scene. In the above scenes, objects such as buildings, trees and roads could be displayed with detailed features, but textures were difficult to be consistent with the real environment. Dias et al. \cite{Dias20063D} used laser scanner to carry out the 3D modeling of real world scenes, which can truly and accurately restore the real world, but it is still not easy in data denoising and texture reconstruction.

The more effective method of generating 3D road network is based on the GIS data, such as satellite imagery and elevation model. 3D scenes in \cite{Thony2016,Agrawal2006,She2017A,Zhi2013} are created by merging vector data and elevation data. In these scenes, roads are displayed in a primitive way, without width, texture and boundary. In the given 3D scenes, GaLin et al. \cite{Galin2010Procedural} discretized the shortest path created by them to generate piecewise clothoid, and then combined with terrain and obstacles so that the roads with slope are created. Wang et al. \cite{Zhiguang20183D} extracted road center axes from simple attribute remote sensing images, and then combined them with terrain to create 3D road network. The roads in the above two scenes both had the width and textures, but did not support the generation and display of intersections. Wu et al. \cite{Wu2015Real} defined a large-scale multi view 3D scenes, which combined 2D road panoramic map, satellite map texture and 3D city model. Li et al. \cite{Yaochen2019Spatiotemporal} reconstructed a 3D scene based on image rendering. They detected the road area from the input image sequence, and then created the road network according to the road boundary control points. However, the roads in \cite{Yaochen2019Spatiotemporal,Wu2015Real} had no height, so that they could not show the changes of slopes. Besides, they did not generate the road network topology data for traffic simulation, either. Bruneton et al. \cite{Eric2008Real} improved the elevation data through combing it with vector data to generate roads with slope fluctuation conforming to the specifications. However, there is still a gap in lanes and texture between the virtual scene and the real world. Besides, the types of intersections supported by Bruneton el al.\cite{Eric2008Real} are also limited.

In short, existing methods can generate 3D realistic scenes, nevertheless, the methods are difficult to automatically generate road details, especially in mountain roads and various kinds of intersections in a large-scale scene generation.

\subsection{Traffic network modeling}
According to the demand of traffic simulation technology, there are many traffic network models to realize traffic simulations and path navigation. The existing traffic simulation software \cite{Yang1996A} mostly uses Node, Link, Segment and Lane to describe the traffic simulation network semantics. These models mainly define the semantic structure of traffic network from the topological data. Wilkie et al. \cite{Wilkie2012Transforming} proposed the arc representation method of Lane vector data, which realized the transformation of rough, low-detailed GIS traffic network data into high-detailed traffic network data. Mao et al. \cite{Tianlu2015An} proposed a method of using compressed point column to represent Lane, which not only keeps the details of traffic network, but also improves the efficiency of obtaining vehicle location information in the process of vehicle motion simulation. Based on the work of \cite{Wilkie2012Transforming} and \cite{Tianlu2015An}, Wang et al. \cite{Wang2014A} proposed a hierarchical traffic network semantic model for vehicle group animation simulation, and the route conflict relation can be generated automatically, which greatly reduces the amount of data input. In the above models, there is a close coupling relation between lanes. If the direction of a few lanes is changed, the system needs to recalculate the topological relationship between all lanes in the traffic network, which is tedious and huge. Moreover, intersection data need to be input, which is difficult to get directly from the existing GIS data.

Inspired by the above traffic models, we introduce a traffic semantic structure description and generate traffic data according to it automatically in this paper.

\begin{figure*}[t]
\centering
\noindent \includegraphics*[width=5.5in, height=2.in, keepaspectratio=false]{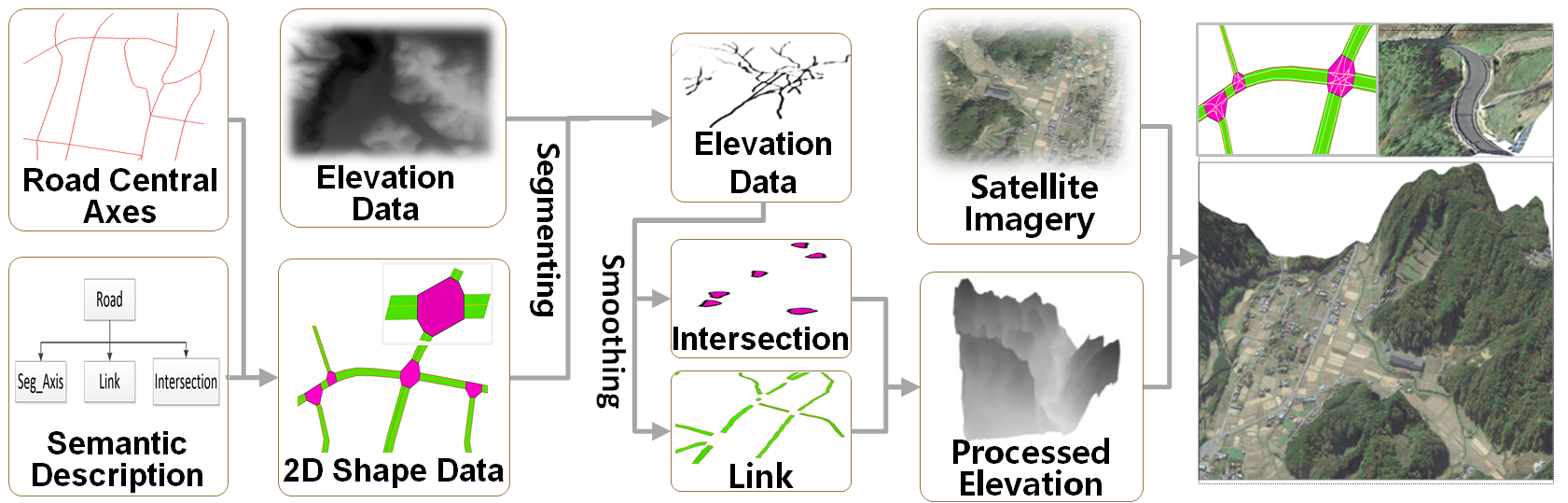}
 \caption{The pipeline of our method. We first define the semantic description of road network, and generate 2D shape data of road network with road center axes as input. Then we use the 2D shape data to segment the original elevation data to get the road area elevation. After the step of smoothing road elevation, we obtain 3D Link and Intersection. Finally, we combine the processed elevation with satellite imagery to reconstruct 3D road networks.}
\label{fig1}
\end{figure*}

\section{Methodology}

Our method generates a realistic 3D road network automatically based on 2D road center axis data, satellite imagery and elevation data. The framework of our method is shown in Fig.\ref{fig1}. There are mainly three processes to generate a 3D road network: semantic descriptions of the road network, semantic data generation in 2D space and 3D road surfaces generation. Firstly, we define the semantic strcture of 2D road network in section 3.1 according to the road center axis data. Then we generate the 2D road shape data in section 3.2. In section 3.3, in order to obtain the elevation data of road area, we segment the elevation data with the help of 2D shape data of Link and Intersection which defined in semantic structure. After that, we smooth the elevation data of road area to get the flat and smooth 3D road surfaces.

\begin{figure}[h]
\centering
\noindent  \includegraphics*[width=3.40in, keepaspectratio=false]{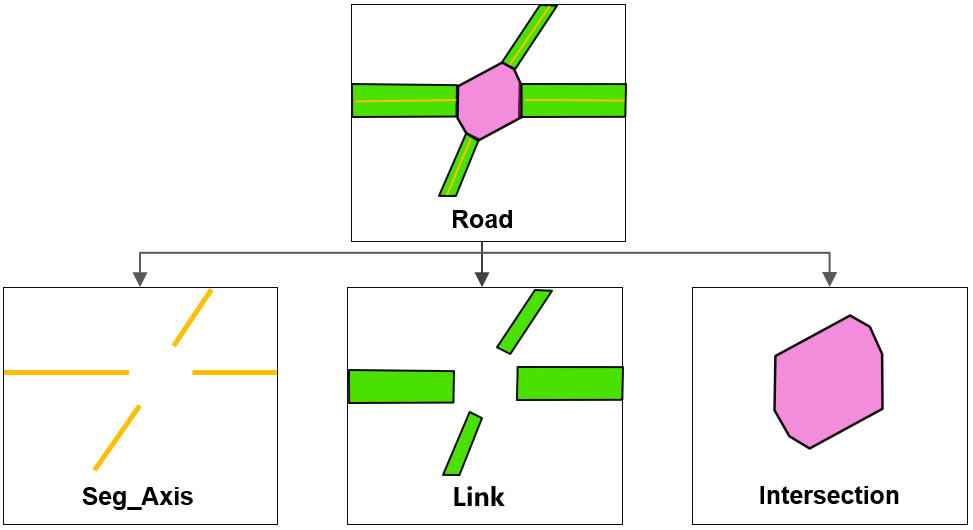}
\caption{The semantic structure of roads.}
\label{fig2}
\end{figure}

\subsection{ Semantic description of the road network}

In order to generate a realistic road network, we introduce a semantic structure of road network (Fig.\ref{fig2}). There are four elements in the semantic structure.

\textbf{Road }is the road network in road scenes. Its topological data is 2D road center axis data. It includes Links and Intersections.

\textbf{Link} is used to describe the half of a road segment. It includes the topological data and geometric data. The topological data contains the connections and adjacency relations between Links. The geometric shape is its boundary. We use curves to describe it.

\textbf{Seg\_Axis }is the center axis of a road segment. It is generated according to the road center axis. It is the junction of two neighboring Links, which describe the left and right half of the road segment. In our method, the Seg\_Axis connects the two neighboring Links. It is easy to prove that one Link has only one Seg\_Axis.

\textbf{Intersection} refers to a road junction. It is the junction of the Links that have connection relations in 3D space. It also includes the topological data and geometric data. Topological data includes the relationship between the Intersection and Links. Geometric data refers to the boundaries of the Intersections. We use a polygon area enclosed by a sequence of points to describe it.

\subsection{Semantic data generation in 2D space}
Semantic data generation in 2D space includes the generation of  the 2D shapes and topological data of all the Links and Intersections in the road network. Here we show the process of how to generate them according to 2D road center axes data and the width of roads. We also generate their topological data in the process.

\begin{figure*}[!t]
\centering
\setlength{\abovecaptionskip}{0.cm}
\noindent \includegraphics*[width=5.54in, height=2.5in, keepaspectratio=false]{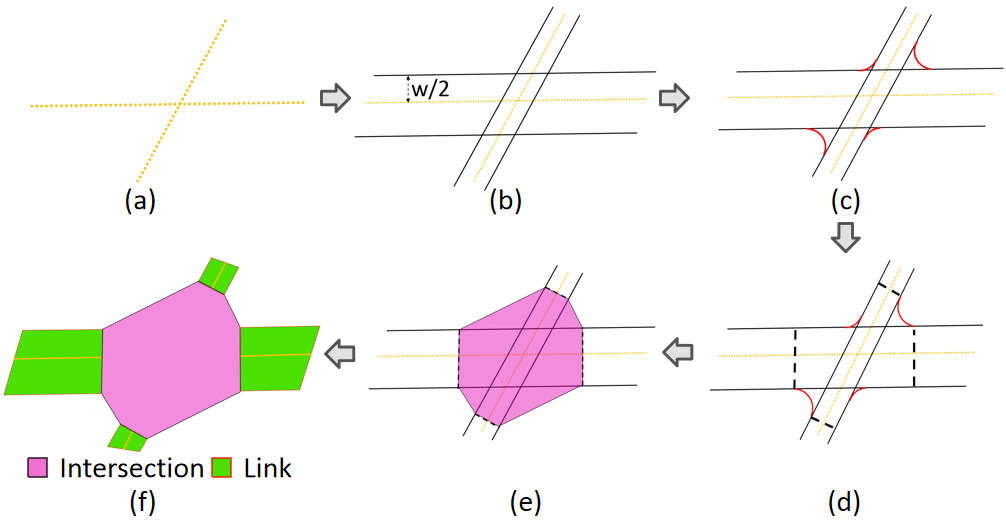}
\caption{The generation process of an Intersection and Links. (a) shows the input road center axes. In (b), we translate the road center axes to get the outer boundaries of roads according to the road width. (c) generates the circular curve. In (d), we generate the tangent lines of road center axes from the end-points of circular curve, and choose the  further one as the boundaries of Intersection. (e) generates the shape of Intersection. (f) generates the shapes of Links. }
\label{fig3}
\end{figure*}

Considering that open source road center axes are usually represented as 2D polyline point sets, which are usually unsmooth, here we firstly use spline interpolations to generate smooth 2D vector data. Then we generate the 2D shapes of all the Links and Intersections of the road network automatically according to the following steps:

Step1: Translating a road center axis along its inner and outer normal directions to generate the two outer boundaries of roads. The distance of the translation is $\frac{w}{2}$. Here \textit{w} is the whole width of the road. Now the road network in a junction becomes several road segments (there are four segments in Fig.\ref{fig3}(b)). 

Step2: Generating the circular curve in the junction of two boundaries. The circular curve is tangent to those two boundaries. Its radius $R$ is defined as follows(red curves in Fig.\ref{fig3}(c)):
\begin{equation}
\centering
{R=\frac{v^2}{127(u+i)} }
\end{equation}

\noindent where $R$ is the radius of the circular curve. $v$ is the design speed of the road, $u$ and $v$ are constant. $u$ represents side-way force coefficient.

Then we collect some points of tangency to generate an Intersection. The point set $P $ in an Intersection must satisfy the following conditions:

1)The circular curve's two points of tangency belong to one Intersection.

2)If $p\mathrm{\in }P$, then $\mathrm{\exists } \ q\mathrm{\in }P\mathrm{,\ and\ }\left|p-q\right|\mathrm{<}L_{dis}\mathrm {}$.  Here {\textbar}$\bullet${\textbar} describes the distance between two points. $L_{dis}$ describes the size of an Intersection. In this paper, we let it be a constant(30 m).

Step3: For each end-point of circular curve, let it be a start-point and generate a line segment between two outer boundaries of the road along the normal of its center axis. If there are two line segments in one road segment, we delete the one which is closer to the boundaries of the Intersection and save the one that further to the boundaries (dotted lines in Fig.\ref{fig3}(d)). Simultaneously, the start point of the deleted line is removed from the Intersection and the end point of the saved line is added to the Intersection.

Step4: Sorting the points in the Intersection with a clockwise rotation and then connecting them to generate a polygon. It is easy to prove that the polygon is a convex polygon. We let it be the boundary of the Intersection (see Fig.\ref{fig3}(e)).

Setp5: Saving the road axes that out of polygons of Intersections and generating the Seg\_Axes. We let the area of a region bounded by a Seg\_Axis, the boundary of a road segment and two edges of polygon of Intersections be a Link. The boundary of road segment is also the Link's boundary, and the Seg\_Axis is also the Link's Seg\_Axis. The generated Link belongs to those Intersections. It is easy to prove that the boundary of the Link is translated, because the Seg\_Axis and the curves of them are parallel.

In order to generate 3D road surfaces, we adopt the curve compression discretization algorithm (Douglas Peucker Algorithm) to discretize and compress the boundary curves of Link, and then get the following descriptions of the geometric shape data of Links and Intersections:\\
\textbf{Link geometric shape data}: It is used to describe the boundary data of a Link. If $A{}_{i}$ is the geometric data of the $i$th Link, it can be expressed as a closed polygon generated from the following point set,
\begin{equation} \label{GrindEQ__2_}
{\rm \{ }q_{i1} {\rm ,}q_{i2} {\rm ,...}q_{iM_{i} } {\rm ,}\tilde{q}_{iN_{i} } {\rm ,}\tilde{q}_{i(N_{i} -1)} {\rm ,...,}\tilde{q}_{i1} {\rm \} }
\end{equation}
where $q_{i1} {\rm ,}q_{i2} {\rm ,...}q_{iM_{i} } $ represents the point column of  the $i$th Link's Seg\_Axis, $\tilde{q}_{i1} {\rm ,}\tilde{q}_{i2} {\rm ,...,}\tilde{q}_{iN_{i} } $ denotes the boundary of the road segment, and $M{}_{i}$ and $N{}_{i}$ are their total number of points, respectively.

\noindent\textbf{Intersection geometric data}: If the current $i$th Intersection aggregates the number of road center axes $k{}_{i}$ , obviously, it can be proved that 2 * $k_i$ Links are generated through the extension of the $k_i$ roads. Then we sort these Links clockwise. We use $n_{ij} \; (j\in [1,2*k_{i} ])$ to represent the $j$th Link, and the geometric data of the current $i$th Intersection can be defined as follows,

\begin{equation} \label{GrindEQ__3_}
{\rm \{ }\tilde{q}_{n_{i1} s_{n_{i1} } } {\rm ,}\tilde{q}_{n_{i2} s_{n_{i2} } } {\rm ,...,}\tilde{q}_{n_{i2k_{i} } s_{n_{i2k_{i} } } } {\rm \} }
\end{equation}

\noindent where $\tilde{q}_{n_{ij} s_{n_{ij} } } \; (j\in [1,2*k_{i} ])$ represents the coordinates of the points in the $j$th Link boundary. If  the road center axis of $j$th Link end at this Intersection, then $s_{n_{ij}}=N_{n_{ij} } $, and $N_{n_{ij} } $ is defined in Eq.2; else if the road center axis start at this Intersection, then $s_{n_{ij}}=1$.

In this paper, we use a symbolic network to define the relation between an Intersection and all of the Links that connect to it. The relation is defined as two types, positive relation and negative relation. For two Links that owns the same Seg\_Axis , if one connects to the Intersection, it is easy to prove that the other one also connects to the same Intersection. We sort them by their boundaries with a clockwise rotation. The relation between the first one and the Intersection is positive, which is represented by "+". The other is a negative relation, which is indicated by "-". This symbolic network description method is suitable for representing the intersecting relation between various types of Intersections and Links.

\subsection{3D road surface generation}
 Here we show how to generate a 3D road surface by combining the above 2D road shape data and elevation data. The elevation data in this paper is a gray-scale image with spatial reference system. The gray value represents the elevation value at the current 2D spatial position. We first segment the elevation data of road areas based on the geometry of Links and Intersections, and then use the segmented areas to smooth road surfaces. After that, a 3D road network with high-detailed road surface is generated.

\subsubsection{Elevation data segmentation}
The elevation image area is represented as $\Omega $, and it is a closed area bounded by the following points,
\begin{equation} \label{GrindEQ__4_}
{\rm \{ }r_{i} {\rm |}i\in [1,4]{\rm \} }
\end{equation}
where $r_{i} {\rm \; (}i\in [1,4])$ represents the vertices of the quadrilateral image. We use $\Omega _{{\rm 1}} $ to define the region enclosed by all Links boundaries and Intersections boundaries obtained in Section 3.2, and it can be expressed as follows ,
\begin{equation} \label{GrindEQ__5_}
\Omega _{1} =\bigcup _{i=1}^{i\le {\rm Link\_ totalnum}}A_{i}  \bigcup _{j=1}^{j{\rm =Intersection\_ totalnum}}B_{j}
\end{equation}

\noindent where Link\_totalnum is the total number of Links and Intersection\_totalnum is the total number of Intersections. $A_{i} \; (i\in [{\rm 1,Link\_ totalnum}])$ and $B_{j} \; (j\in [1,{\rm Intersection\_ totalnum}])$  represent the boundary of the i-th Link and j-th Intersection, respectively (defined in Section 3.1). Then we split the $\Omega $ based on the $\Omega _{1} $ , and $\Omega $ can be defined a following equation,
\begin{equation} \label{GrindEQ__6_}
\Omega =\Omega _{1} \bigcup \bar{\Omega }_{1}
\end{equation}
where $\bar{\Omega }_{1} $ represents the complement set of $\Omega _{1} $. Obviously, it can be proved that $\Omega _{1} $ represents the elevation data of road areas in the scene, and any sub area of $\Omega _{1} $ is the closed polygon area enclosed by the boundary of Intersection or the Link.

\subsubsection{Smooth processing of road surfaces}

This part introduces the approach of smoothing the road surface area $\Omega _{1} $. For the elevation value \textit{z} at any position (including the boundary) in the closed polygon area $B_{i} \; (i\in [1,{\rm Intersection\_ totalnum}])$, it is equal to the average value of the elevation value of all the vertexes on the Intersection boundary. Then, $\forall p(x,y,z)\in B_{i} $,

\begin{equation} \label{GrindEQ__7_}
z=\frac{1}{2*k_{i} } \sum _{n=1}^{n=2*k_{i} }h_{in}
\end{equation}

\noindent where h${}_{in(}$$n\in 2*k_{i} $${}_{)}$ is the elevation value of the n-th vertex in the i-th Intersection(\textit{k${}_{i}$} is defined in 3.2).

As for the closed polygon area $A_{i} \; (i\in [{\rm 1,Link\_ totalnum}])$, if the elevation of each point in the interior $A_{i} \; $ is assigned directly according to the download elevation data, the road surfaces would be uneven. In order to avoid this problem, we smooth the elevation of the road center axes firstly, and then smooth the elevation of other positions in the $A_{i} \; $ based on the road center axes.

\textbf{Smoothness of road center axis}:  The first step is the curve fitting of road center axis in the 2D Cartesian coordinate space. Then the curve fitting is carried out in the Frenet frame space which is formed along the fitting curve direction and road network elevation direction to obtain the road profile curves.

In the 2D Cartesian coordinate space, after the curve fitting of the polyline points column $q_{i1} {\rm ,}q_{i2} {\rm ,...}q_{iM_{i} } $ of $A_{i} \; $,  the arc length parameterized curve $C_{i} $ is obtained,
\begin{equation} \label{GrindEQ__8_}
C_{i} =C_{i} (s)=(x(s),y(s))
\end{equation}

$s$ is the arc length. $(x,y)$ is the 2D coordinate of any point in the curve $C_{i} $. In that way, for any arc length \textit{s} on the curve, it is easy to calculate the elevation value $h(s)$ of this position $h(s)=h(x(s),y(s))$.

For each point on the 2D curve $C_{i} $, the coordinate space is constructed with the tangent direction and the elevation direction of the road network (the direction of the external normal of the plane where the curve is located) as the coordinate axes. The coordinate space is a 2D Frenet frame space. We generate the profile curves in the Frenet frame space. According to the road design and construction standards, the profile curves of a road are the curves with monotone and constrained curvature: the curvature is less than 0.1. Here we first generate control vectors according to the above constraint in the space and then use the piecewise B-spline curve with monotone curvature to generate the profile curves.

1)Control vector generation:

Sampling points by equal distance $\Delta s$ along the curve $C_{i} $. According to the civil engineering restrictions, the slope of road usually does not exceed 0.08\%. Therefore, if the absolute value of the curvature difference between two vector points is more than 0.1, the later vector will not be collected in the control vector set. Then we can get the following control vector set:

\noindent
\[(0,h(0)),\; ({\rm n}_{1} \Delta s,h({\rm n}_{1} \Delta s)),......({\rm n}_{{\rm Total}} \Delta s,h({\rm n}_{{\rm Total}} \Delta s)))\]
where ${\rm n}_{{\rm Total}} \le \left\lfloor S/\Delta s\right\rfloor $, $S$ is the total arc length of the curve $C_{i} $, and $\left\lfloor \bullet \right\rfloor $ is the rounding symbol.

2)Profile curves generation

Here we use the piecewise B-spline curve with monotone curvature to generate the profile curves. Each piece of the cubic B-spline curve is generated by the method presented by Wang et al.[6]. We fit the above point column to get the relation curve $\tilde{C}_{i} $ between \textit{s} and \textit{h},
\begin{equation} \label{GrindEQ__9_}
\tilde{C}_{i} =h(s)
\end{equation}
according to the Eq.9, the elevation values of each coordinate point in $q_{i1} {\rm ,}q_{i2} {\rm ,...}q_{iM_{i} } $ are calculated.

\textbf{Smoothness within the Link geometric area}: According to Section 3.1, the Link area is a closed area derived from the extension of road axis. In the Link area, the elevation value at any point is equal to its nearest point of road axis, which means that the elevation value in the direction of the road section profile is constant.

\begin{figure}[h]
\centering
\noindent \includegraphics*[width=3.4in,height=3.05in, keepaspectratio=false]{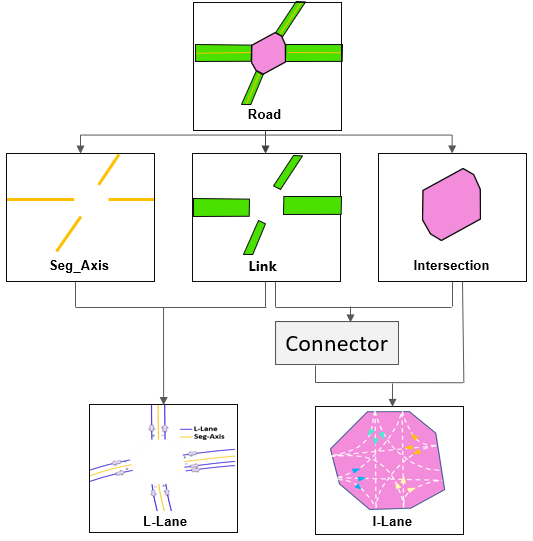}
\caption{The road structure for traffic simulations}
\label{fig4}
\end{figure}

\section{Implementation and application}
In this section, we introduce the implementation of our method. Besides, we also show how to generate the road structure which can be used for traffic simulation directly based on the semantic structure of 3D scenes defined in 3.1.

\begin{figure}
\centering
\setlength{\abovecaptionskip}{0.cm}
\noindent \includegraphics*[width=2.55in, height=1.05in, keepaspectratio=false]{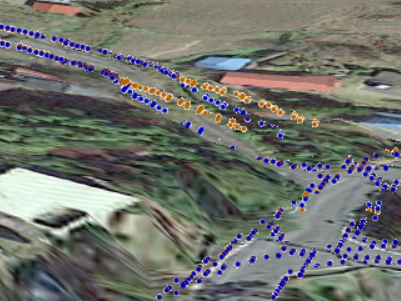}
\includegraphics*[width=2.55in, height=1.05in, keepaspectratio=false]{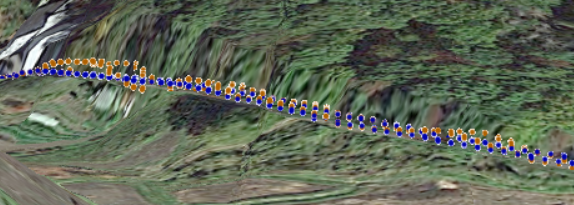}
\noindent \includegraphics*[width=2.55in, height=1.05in, keepaspectratio=false]{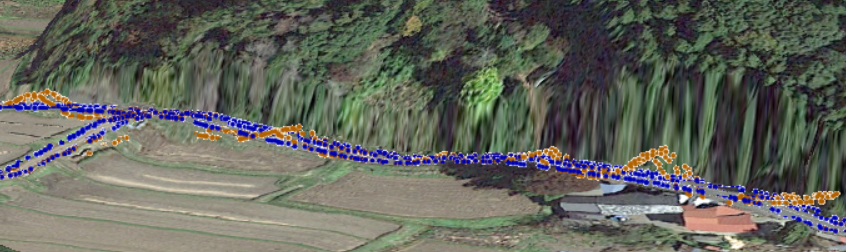}
\includegraphics*[width=2.55in, height=1.05in, keepaspectratio=false]{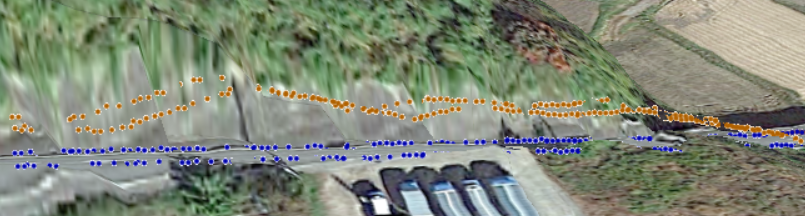}
 \caption{The results of our approach with the process of the smoothness of road center axes (blue points) and without (orange points).}
\label{fig5}
\end{figure}

We first generate 2D shape data of road with the road center axis data as input according to the road semantic structure. The roads are made up of three elements, Link, Intersection and Seg\_Axis. Then we segment the original elevation data by using the generated 2D road shape, and get the elevation data with only road area. Every step is mutually linked and essential. After that, smoothing the elevation data of road is our key step. Here we smooth the cross section of the road and the longitudinal section of the road respectively. The orange points in Fig.\ref{fig5} are collected from the output experimental results under the condition that only the road cross section is flat. At this time, the road is tortuous, not in line with the reality. The blue points are collected from the output experimental results after the smoothness of road center axis. It can be seen that the road is handled better. At last, we input the satellite imagery as texture to reconstruct the 3D scenes.

Two elements (Lane and Connector) are added (Fig.\ref{fig4}) in traffic semantic structure.
\textbf{Lane} describes a trajectory of vehicles along roads. There are two types according to the data generation:

\textbf{L\_Lane} describes a trajectory of vehicles in a Link. In this paper, the geometric shape data of L\_Lane is generated by extending Seg\_Axis of the Link, and the extension distance is determined by the width of the lane. Obviously, a L\_Lane starts from an Intersection and ends to another Intersection. Here we also use ``+'' and ``-'' to describe the relation between a L\_Lane and an Intersecion. If the L\_Lane starts from the Intersection, the relation is ``-'', otherwise, it is ``+''.

\textbf{I\_Lane} represents a trajectory of vehicles in an Intersection. Its topological data defines as follows:

\begin{itemize}
\vspace{-0.3cm}
\item It starts from the end point of a L\_Lane that has the``+'' relation with the Intersection, and it also tangents to the L\_Lane at that end point;
\vspace{-0.3cm}
\item It ends to the starting point of ``-'' L\_Lane that has the``-'' relation with the Intersection , and it also tangents to the L\_Lane at that starting point;
\vspace{-0.3cm}
\item Each point on the I\_Lane has second-order smoothness;
\vspace{-0.3cm}
\item There is only one I\_Lane between an end point of one L\_Lane and a starting point of another L\_Lane;
\end{itemize}

\begin{algorithm}[h]
\caption{The algorithm for I\_Lane generation}
\begin{algorithmic}
\label {Algorithm1}
\STATE //The number of  L\_Lanes that have "+"relation with the Intersection
\STATE int \emph{fromN};
\STATE //The number of L\_Lanes that have "-" relation with the Intersection
\STATE int \emph{toN};
\STATE //$M$ is the number of Links that have relations with the Intersection
\FOR{$(i=1; i\mathrm{<}M; i=i+2)$}
\STATE $\!k=i+1;$
\STATE $\!num=1;$
\WHILE {$\!(num\mathrm{<}=M/2)$}
\IF{$k>M $}
\STATE $k=k-m;$
\ENDIF
\STATE $fromN=num; toN=num;$
\IF{$(fromN\mathrm{>}N1:\texttt{total number of Lane in Link1})$}
\STATE $fromN=N1;$
\ENDIF
\IF{$(ToN\mathrm{>}N2:\texttt{total number of Lane in Link2})$}
\STATE $toN=N2;$
\ENDIF
\STATE Generate one I\_Lane=Bezier (Vectors of the end point of the $fromN$th L\_Lane in the $i$th Link and of the start point of the to $N$th L\_Lane in the $k$th Link);
\IF{$(k!=i+1)$}
\STATE $num++;$
\ENDIF
\STATE $k++;$
\ENDWHILE
\ENDFOR
\end{algorithmic}
\end{algorithm}

\vspace{-0.4cm}
\begin{figure}[h]
\centering
\setlength{\abovecaptionskip}{0.cm}
\noindent \includegraphics*[width=2.4in, height=2.4in, keepaspectratio=false]{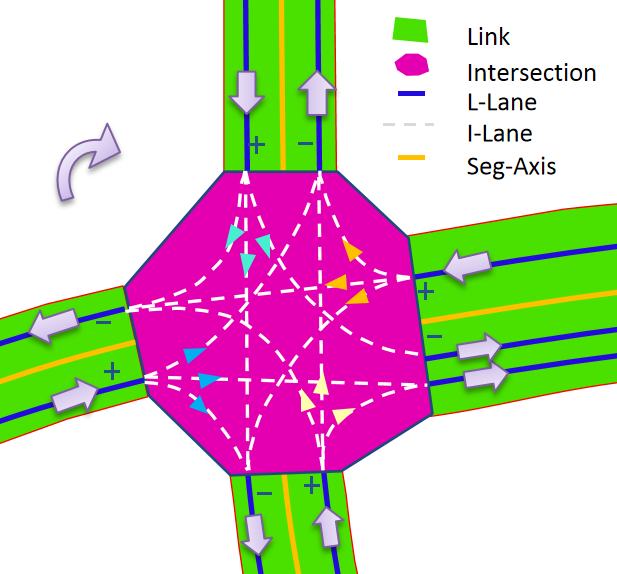}
\caption{ The road structure  for traffic simulation. The yellow lines represent Seg\_Axis. Blue lines with purple arrowhead are the L\_Lanes extended from Seg\_Axis. The white lines are the I\_Lanes that start from ``+''L\_Lanes and end at ``-''L\_Lanes.}
\label{fig6}
\end{figure}

According to the definition of I\_Lane and the actual situations, it is not difficult to generate its geometric data automatically. The generation algorithm is shown in Algorithm \ref{Algorithm1}.

\begin{figure}
\centering
\setlength{\abovecaptionskip}{0.cm}
\noindent \includegraphics*[width=1.45in, height=1.45in, keepaspectratio=false]{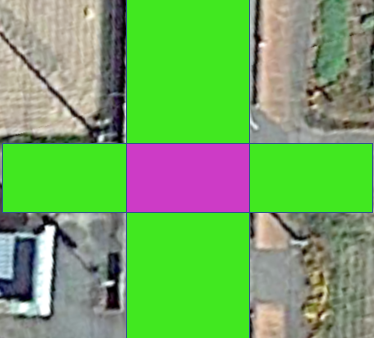}(a)
\noindent \includegraphics*[width=1.45in, height=1.45in, keepaspectratio=false]{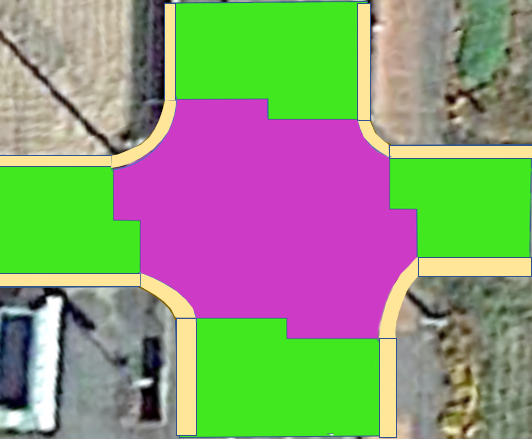}(b)
\vspace{+0.18cm}

\noindent \includegraphics*[width=1.45in, height=1.45in, keepaspectratio=false]{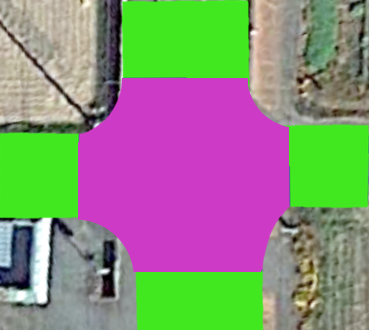}(c)
\noindent \includegraphics*[width=1.45in, height=1.45in, keepaspectratio=false]{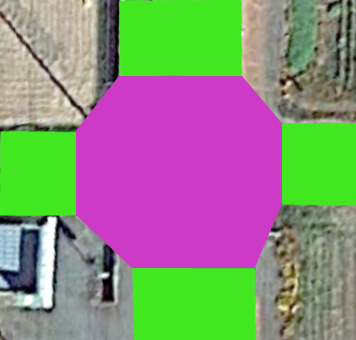}(d)

\vspace{+0.18cm}

\caption{The comparison of intersection modeling. (a) shows the Intersection generated by CityEngine\cite{CityEngine2016} and Brunton\cite{Eric2008Real}. (b) illustrates the Intersection defined by Nguyn\cite{Ho2016Realistic}. (c) represents the Intersection modeled in Chen\cite{Chen:Street:2008} and Cura\cite{StreetGen2018}. The Intersection in (d) is generated by our methods. }
\label{fig7}
\end{figure}

\textbf{Connector} is the connection between two Links. It only has topological data. If there is an I\_Lane which connects one L\_Lane in Link1 to another L\_lane in Link2, then Link1 is connected to Link2.

\begin{table*}[htb]
\caption{\label{tab1}The feature comparisons between some road network models and our method}
\label{tab1}
\vspace{+0.2cm}
\centering
\begin{tabular}{p{0.8in}p{0.8in}p{1.4in}p{1.2in}p{1.2in}} 
\toprule
\textbf{Scenes} & \textbf{Dimensions of Scene} & \textbf{Sense of reality} & \textbf{Generation of Intersection } & \textbf{Generation of Traffic data} \\
\midrule
Bruneton\cite{Eric2008Real} & 3D & Virtual simulation &  Semi-automatic & Not support\\ 
Cura\cite{StreetGen2018} & 2D & Realistic reconstruction  & Automatic & Automatic \\
Galin\cite{Galin2010Procedural} & 3D & Virtual simulation & Not support & Not support \\
Wang\cite{Zhiguang20183D}& 3D & Realistic reconstruction & Not support & Not support\\ 
Our method & 3D &  Realistic reconstruction & Automatic & Automatic \\
\bottomrule
\end{tabular}
\end{table*}

\section{Results}

This section verifies the effectiveness and efficiency of generating a large-scale 3D road network automatically by the method proposed in this paper. Firstly, the comparison between our method and some typical methods for road networks in recent years is given. After that, the calculation efficiency is analyzed in detailed. At last, we show some typical scenes generated by our method. Results showed that our method can generate large-scale 3D scenes with high-detailed road surfaces for traffic simulation when users only have satellite imagery data, elevation data and road center axes.

\subsection{Comparison study}

\begin{figure*}[!t]
\centering
\setlength{\fboxrule}{1pt}
\setlength{\fboxsep}{0.5mm}
\fbox{\includegraphics*[width=1.28in, height=1.30in, keepaspectratio=false,trim=0.00in 0.06in 0.00in 0.00in]{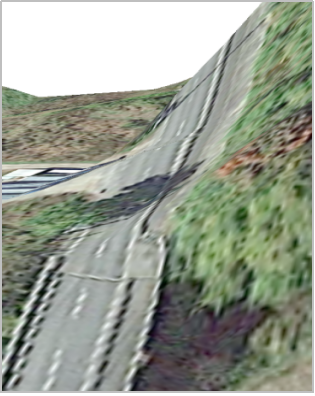}}
\fbox{\includegraphics*[width=1.28in, height=1.30in, keepaspectratio=false]{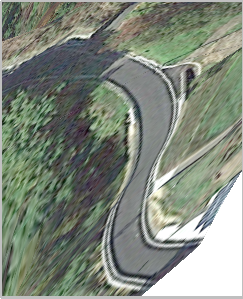}}
\fbox{\includegraphics*[width=1.28in, height=1.30in, keepaspectratio=false]{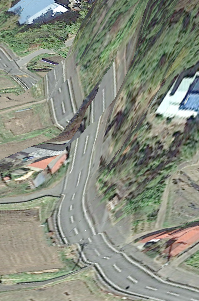}}
\fbox{\includegraphics*[width=1.28in, height=1.30in, keepaspectratio=false]{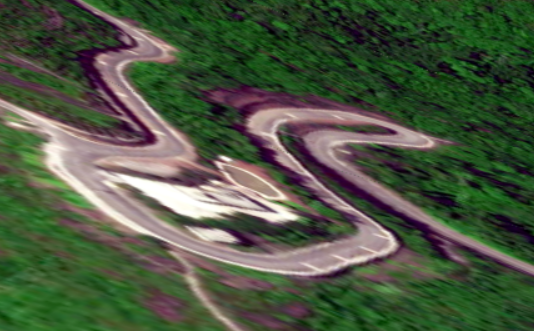}}
\vspace{+0.08cm}
\\

\fbox{\includegraphics*[width=1.28in, height=1.30in, keepaspectratio=false]{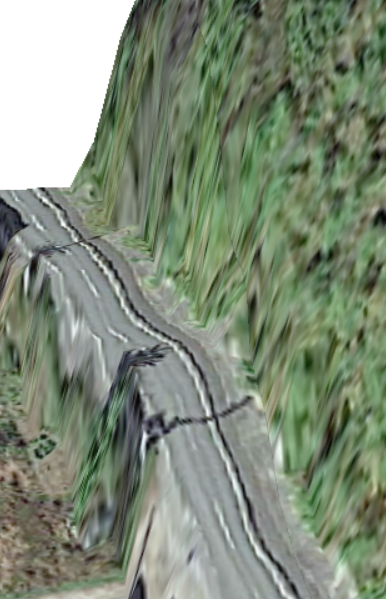}}
\fbox{\includegraphics*[width=1.28in, height=1.30in, keepaspectratio=false]{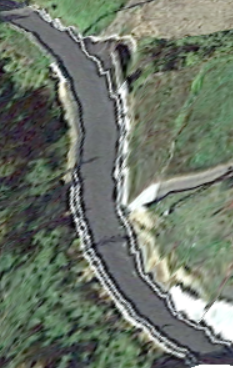}}
\fbox{\includegraphics*[width=1.28in, height=1.30in, keepaspectratio=false]{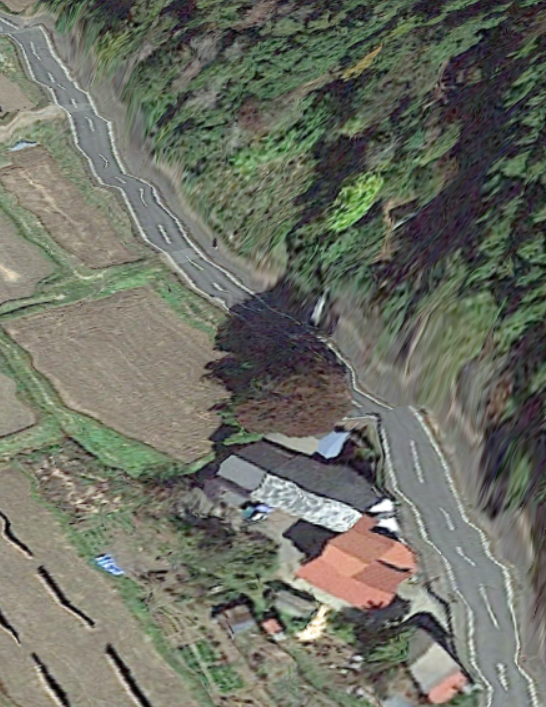}}
\fbox{\includegraphics*[width=1.28in, height=1.30in, keepaspectratio=false]{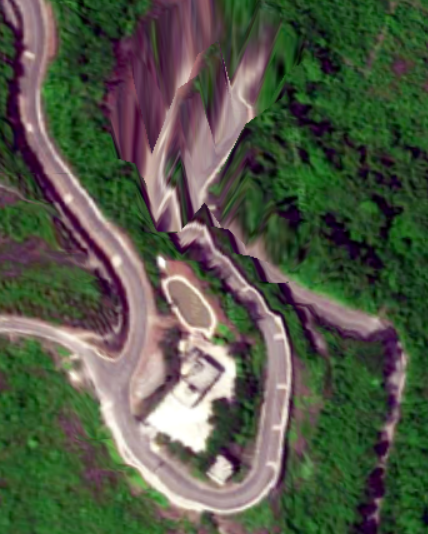}}
\vspace{+0.08cm}
\\

\fbox{\includegraphics*[width=1.28in, height=1.30in, keepaspectratio=false]{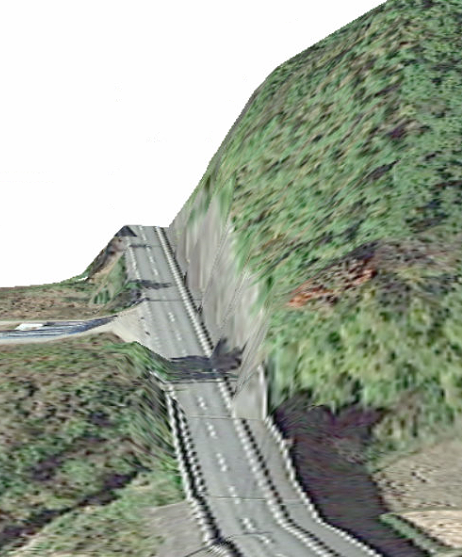}}
\fbox{\includegraphics*[width=1.28in, height=1.30in, keepaspectratio=false]{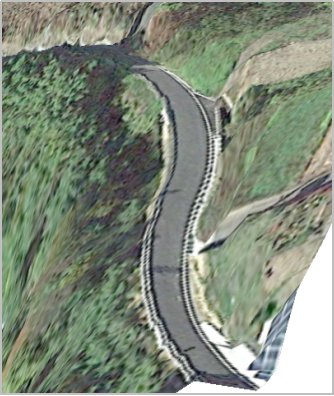}}
\fbox{\includegraphics*[width=1.28in, height=1.30in, keepaspectratio=false]{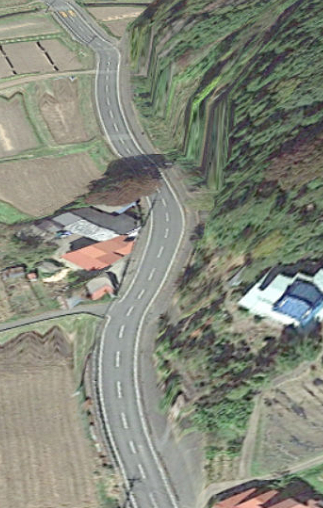}}
\fbox{\includegraphics*[width=1.28in, height=1.30in, keepaspectratio=false]{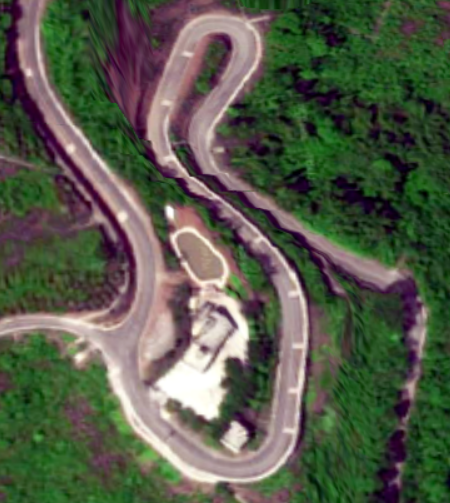}}
\vspace{+0.08cm}
\\

\fbox{\includegraphics*[width=1.28in, height=1.30in, keepaspectratio=false]{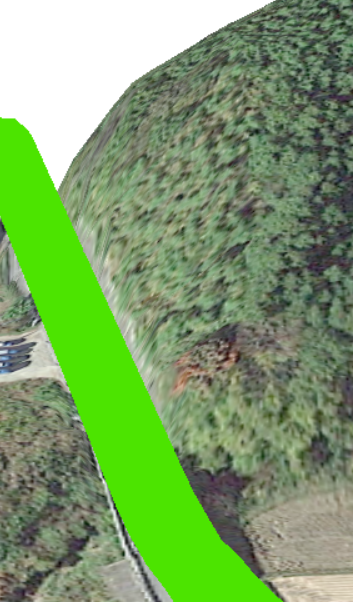}}
\fbox{\includegraphics*[width=1.28in, height=1.30in, keepaspectratio=false]{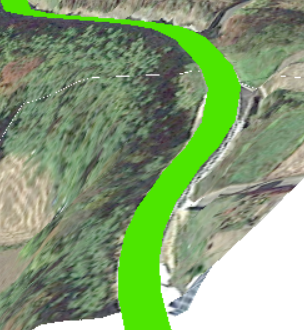}}
\fbox{\includegraphics*[width=1.28in, height=1.30in, keepaspectratio=false]{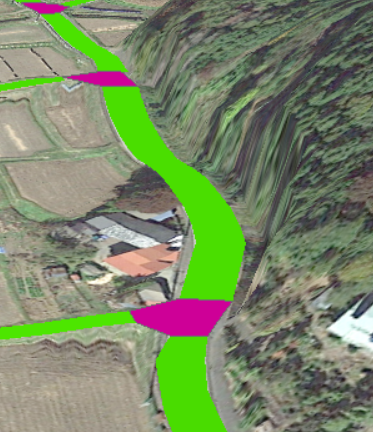}}
\fbox{\includegraphics*[width=1.28in, height=1.30in, keepaspectratio=false]{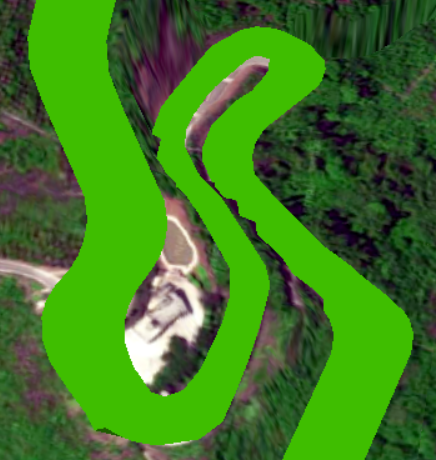}}

\caption{ The first row shows the 3D road surfaces generated by the method of texture mapping based projection using ArcGIS. The second row figures are generated by Wang\cite{Zhiguang20183D}. The rest of these fs are generated by our method. The curvature of these three roads increases in turn from left to right. The bottom figures show the link and intersection we generated.}
\label{fig10}
\end{figure*}

Our method has significant advantages in automatic generation of 3D road networks especially intersections. Detailed feature comparisons between some typical methods and our method are shown in Table \ref{tab1}. Comparing some methods in showing a virtual simulation for real road networks and generating complex intersections, our method can not only generate a realistic 3D road network reconstructions based on measurement data, but also generate complex  intersections and traffic data automatically.  There are two kinds of scenes. The scenes in \cite{Eric2008Real,Galin2010Procedural} are created by virtual simulation, the other scenes \cite{StreetGen2018,Zhiguang20183D} are generated based on the real data completely and we classified it as realistic reconstruction. Galin\cite{Galin2010Procedural} and Wang\cite{Zhiguang20183D} supported for generating neither intersections nor traffic data. Brunton\cite{Eric2008Real} could generate simple trigeminal intersection and the four corners, but they did not mention complex intersections. As for the scene in \cite{StreetGen2018}, it is 2D scene, so it is limited to show the slope and flatness of roads. We chose the regions in Japan and Chongqing for the experiment, and compared the results of 3D intersections and roads in Fig.\ref{fig10} and Fig.\ref{fig11}.

\textbf{ Generation of intersection:} Here we compare several definations of intersections. Intersection is modelled as a square connected to neighboring rectangles with narrow triangles in CityEngine and Bruneton\cite{Eric2008Real} (see Fig.\ref{fig7}(a). The intersection in Nguyn\cite{Ho2016Realistic} is split into non-overlapping zones: traveled way and shoulder ( see Fig.\ref{fig7}(b)), but the intersections formed by high-curvature and high-speed trajectories are ignored.  Chen\cite{Chen:Street:2008} generates intersections by using specification templates defined in \cite{2000Modelling}, but they only have X-type and T-type intersections. Cura\cite{StreetGen2018} also generates intersections by combining the crossing angle of roads, and this method is suitable for intersections with various shapes (see Fig.\ref{fig7}(c)). However, when it is used in country roads, there is often a altitude difference between the road area and the non road area, which leads to the stretch of the road surface area. So we designed a slightly wider intersection to keep the intersection flat, our model of intersections is shown in Fig.\ref{fig7}(d).

\textbf{3D road network:} ArcGIS allows users to combine vector maps with image based information, like aerial photographic data, and to project them onto a terrain height field surface model. The quality of 3D roads is completely depends on the quality of elevation data ( see the figures of first row in Fig.\ref{fig10} ). Wang\cite{Zhiguang20183D} also noticed this problem. They considered the height of road cross section. We tested this method in urban area and country area, and we find that it performs well in urban area. However, road longitudinal section is still not smooth enough in country area (see second row figures of Fig.\ref{fig10}).  As for the 3D scenes based on 3D models, the height of roads is constant. It's rare to see a road with a rolling slope. Compared with these methods, our method ensures the smoothness of both cross section and longitudinal section (see the figures of third row in Fig.\ref{fig10}).

\begin{figure}[!t]
\centering
\setlength{\abovecaptionskip}{0.cm}
\noindent \includegraphics*[width=2.64in, height=1.80in, keepaspectratio=false]{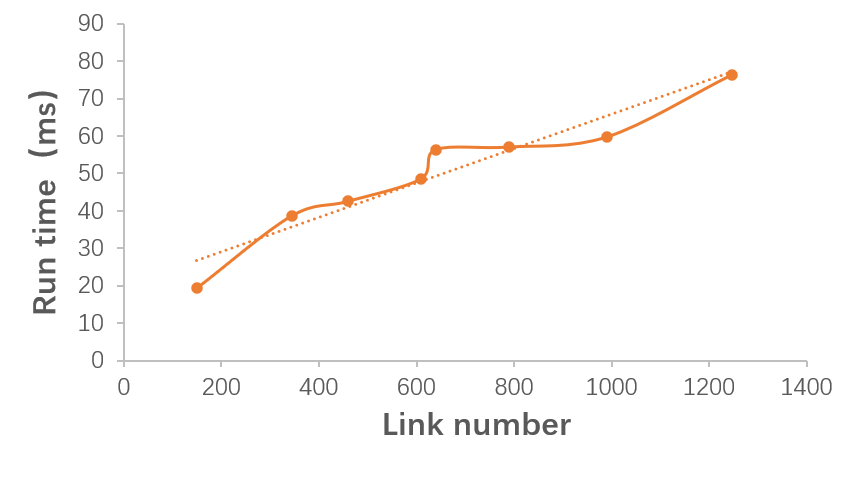}
\caption{The relation between Link number and run time}
\vspace{-0.5cm}
\label{fig9}
\end{figure}

\subsection{Performance analysis}

After the 3D road network is generated by this method, the semantic data of traffic  network is generated automatically. The efficiency of generating semantic data of traffic network is analyzed as follows. Our experiments were implemented on the desktop PC with an Intel Xeon E5-1620 CPU and two graphics cards (NVIDIA GTX 1080 Ti).

We first generate 2D shape data of road network within 10 kilometers. Fig.\ref{fig9} shows the relation between the number of Link in the road network and the time used in calculation. It can be seen that the calculation time increases as the number of road center axes increases.

\begin{figure*}[htb]
\centering
\includegraphics*[width=6.04in, height=3.03in, keepaspectratio=false]{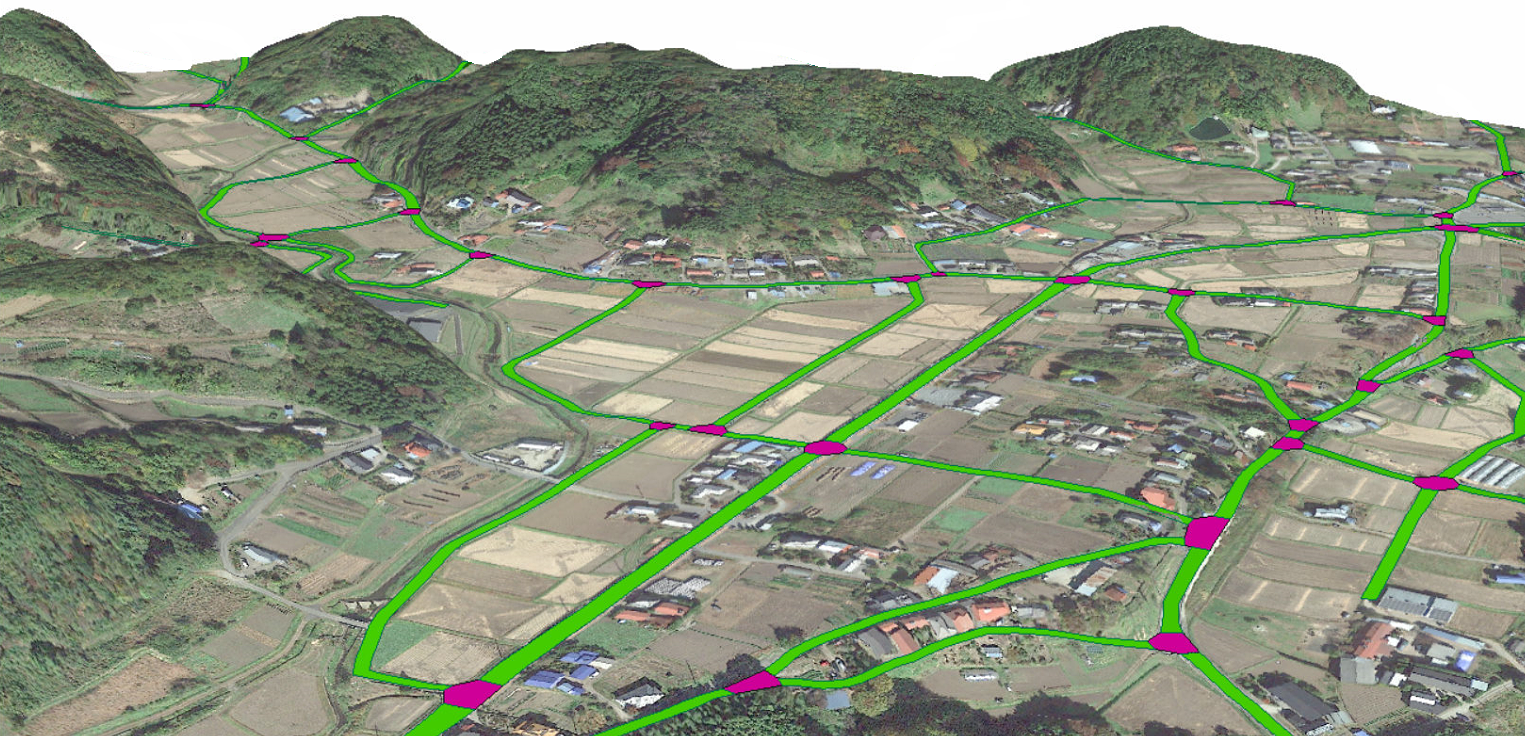}
\caption{ The 3D road network generated by our method.}
\label{fig8}
\end{figure*}

\subsection{Validation for 3D road scenes reconstruction}
\vspace{-0.029cm}
In order to demonstrate the effect of our method, we use it to reconstruct a large-scale 3D rural scene, including the reconstruction of Mountain roads, ramps, various types of Intersections, etc. The result is shown in Fig.\ref{fig8}.

\textbf{Mountain road}: Fig.10 shows several types of mountain road. The top three subfigures are the results using texture mapping based projection. The bottom three subfigures are the results using our method. The degree of road curvature increases from left to right. It shows that the 3D road surfaces generated by our method are in line with actual road specifications, even though the road is rugged.

\textbf{Intersection}: The 3D intersections generated by our method are also coincide with the actual environment. As shown in Fig.11, our road surfaces of intersections are more flat than before, no matter what kind of shape it is.

\begin{figure*}[!t]
\centering
\setlength{\fboxrule}{1pt}
\setlength{\fboxsep}{0.5mm}
\fbox{\includegraphics*[width=0.75in, height=1.05in, keepaspectratio=false]{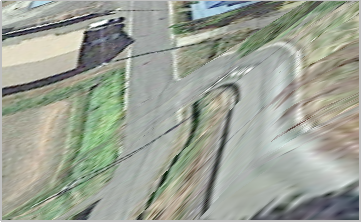}}
\fbox{\includegraphics*[width=0.75in, height=1.05in, keepaspectratio=false]{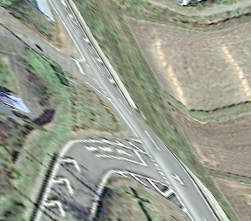}}
\fbox{\includegraphics*[width=0.75in, height=1.05in, keepaspectratio=false]{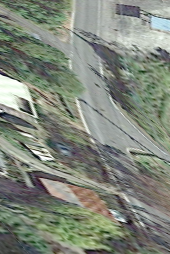}}
\fbox{\includegraphics*[width=0.75in, height=1.05in, keepaspectratio=false]{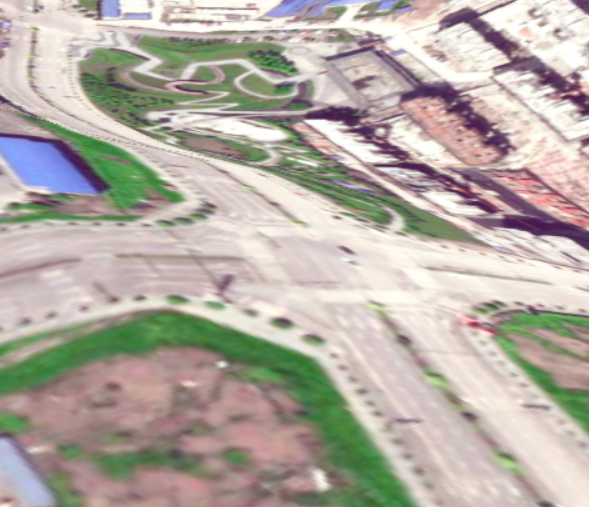}}
\fbox{\includegraphics*[width=0.75in, height=1.05in, keepaspectratio=false]{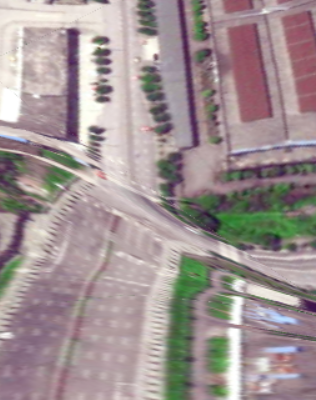}}
\fbox{\includegraphics*[width=0.75in, height=1.05in, keepaspectratio=false]{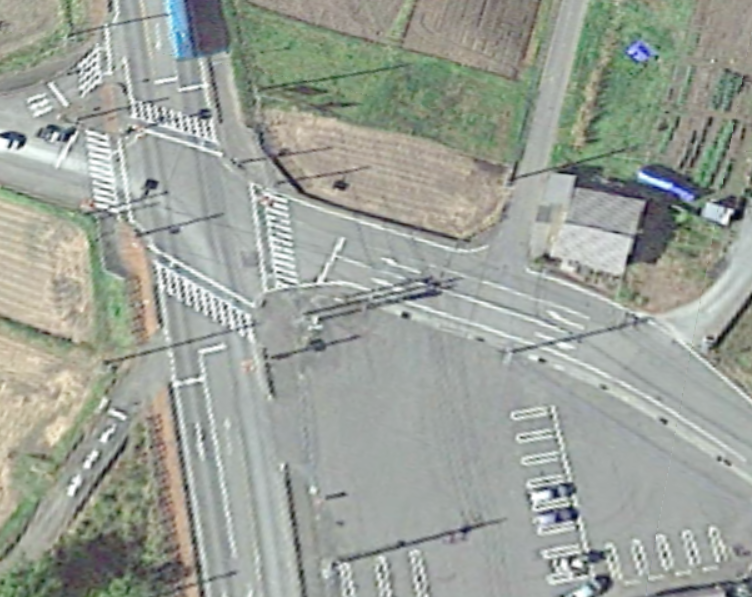}}

\vspace{+0.08cm}
\fbox{\includegraphics*[width=0.75in, height=1.05in, keepaspectratio=false]{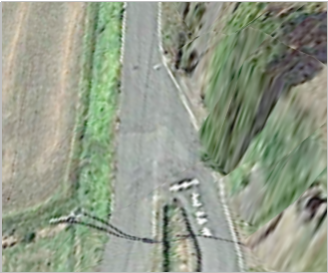}}
\fbox{\includegraphics*[width=0.75in, height=1.05in, keepaspectratio=false]{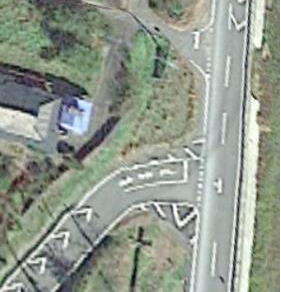}}
\fbox{\includegraphics*[width=0.75in, height=1.05in, keepaspectratio=false]{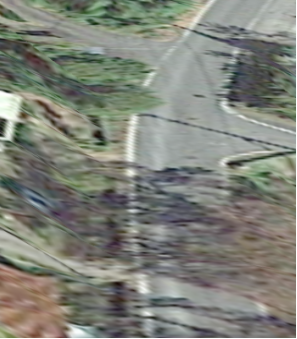}}
\fbox{\includegraphics*[width=0.75in, height=1.05in, keepaspectratio=false]{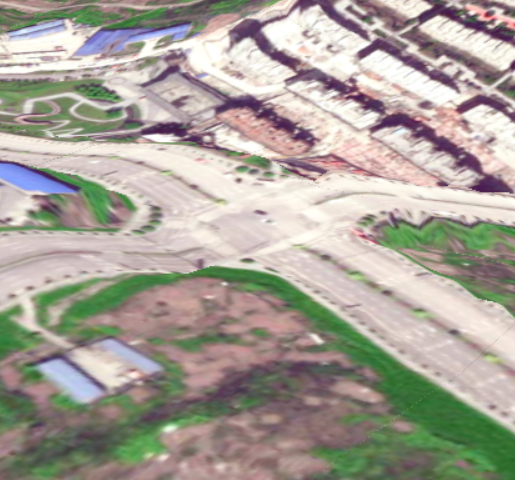}}
\fbox{\includegraphics*[width=0.75in, height=1.05in, keepaspectratio=false]{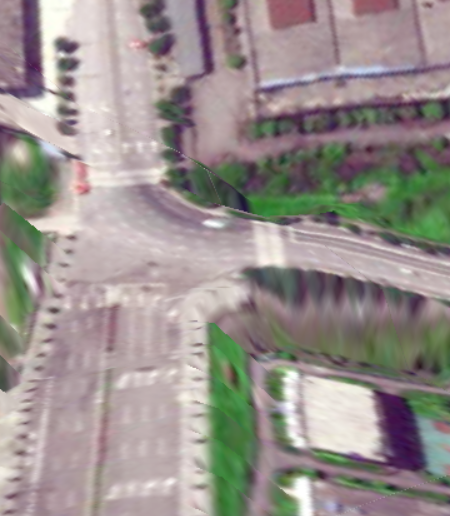}}
\fbox{\includegraphics*[width=0.75in, height=1.05in, keepaspectratio=false]{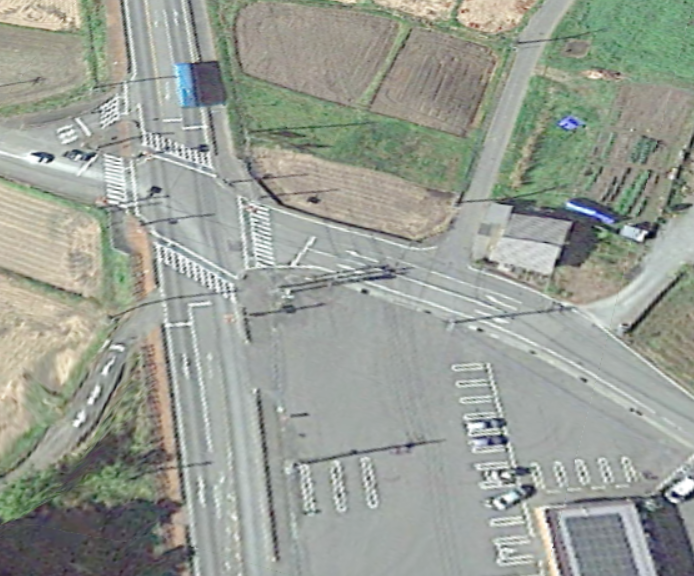}}

\vspace{+0.08cm}
\fbox{\includegraphics*[width=0.75in, height=1.05in, keepaspectratio=false]{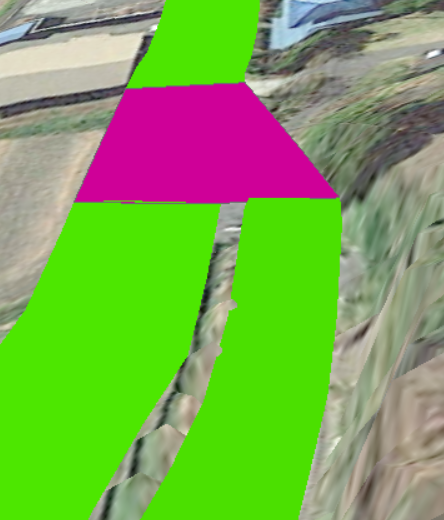}}
\fbox{\includegraphics*[width=0.75in, height=1.05in, keepaspectratio=false]{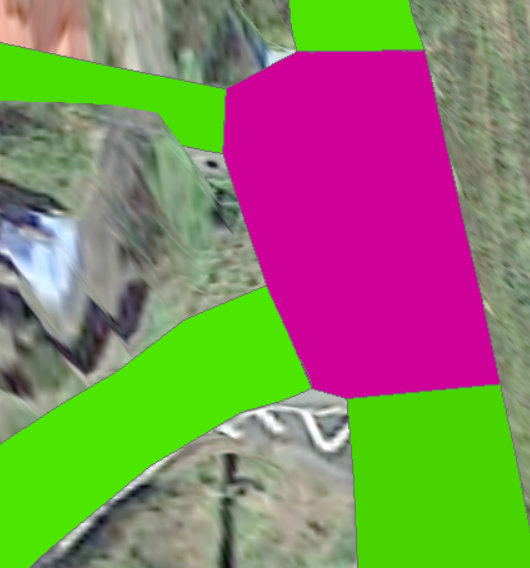}}
\fbox{\includegraphics*[width=0.75in, height=1.05in, keepaspectratio=false]{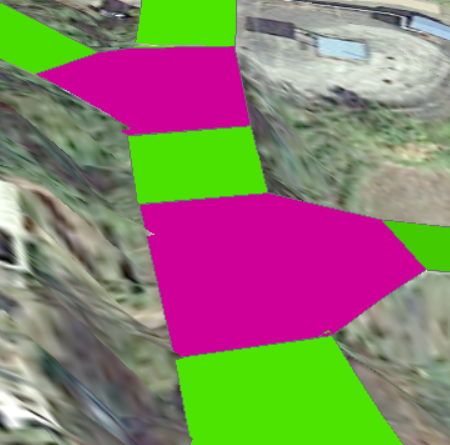}}
\fbox{\includegraphics*[width=0.75in, height=1.05in, keepaspectratio=false]{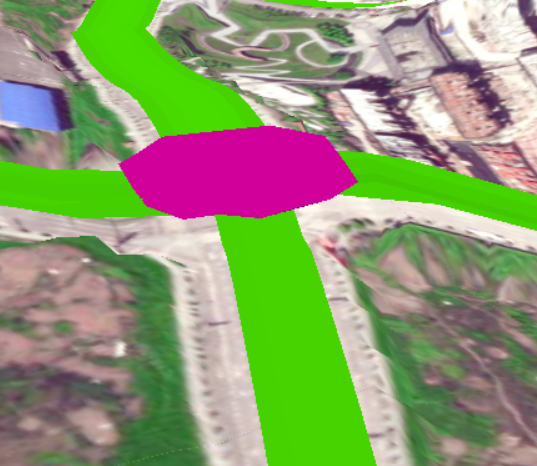}}
\fbox{\includegraphics*[width=0.75in, height=1.05in, keepaspectratio=false]{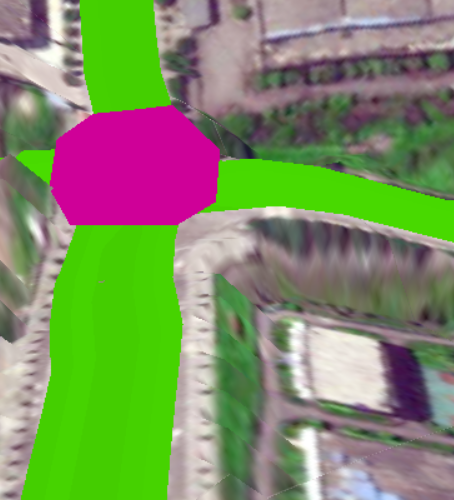}}
\fbox{\includegraphics*[width=0.75in, height=1.05in, keepaspectratio=false]{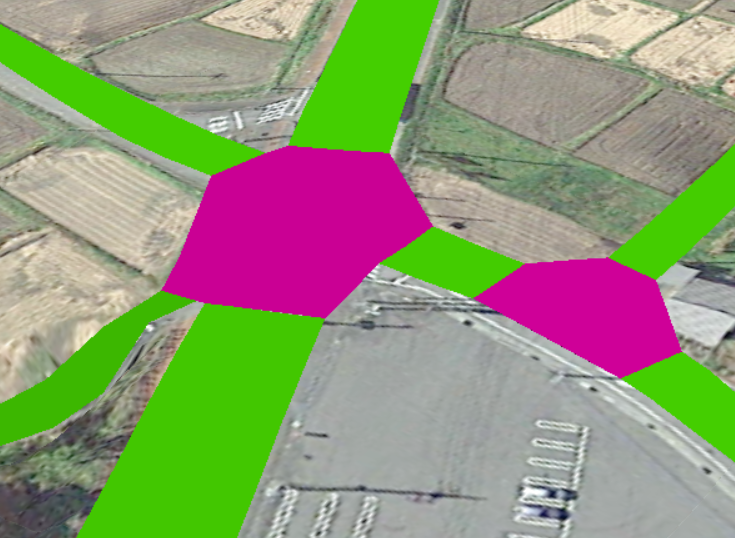}}

\caption {The comparison of different kinds of 3D Intersections between original data(top) and processed data(middle). The bottom figures show the link and intersection we generated.}
\label{fig11}

\end{figure*}

\subsection{Validation for traffic simulation application}

\begin{figure*}[!t]
\centering
 \includegraphics*[width=1.20in, height=1.20in, keepaspectratio=false]{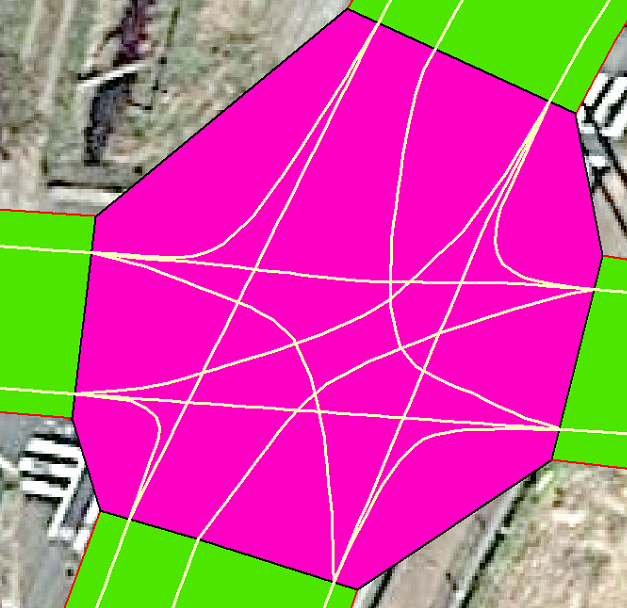}
 \includegraphics*[width=1.20in, height=1.20in, keepaspectratio=false]{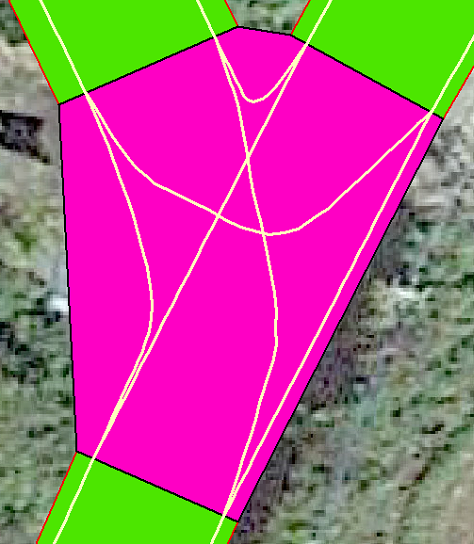}
 \includegraphics*[width=1.20in, height=1.20in, keepaspectratio=false]{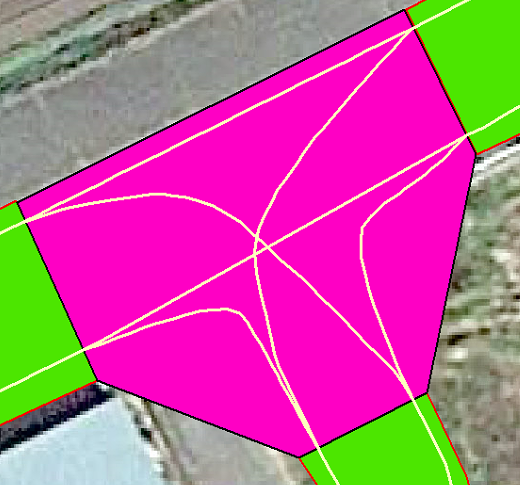}
 \includegraphics*[width=1.20in, height=1.20in, keepaspectratio=false]{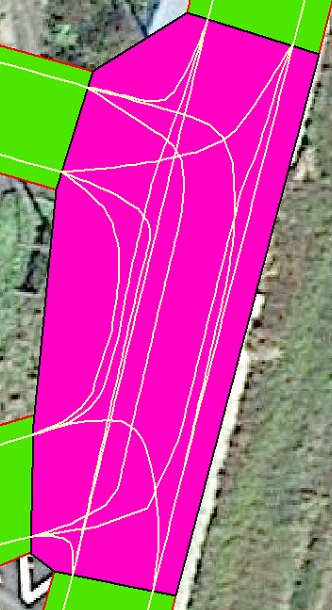}
\caption{I\_Lanes and L\_Lanes in different kinds of Intersections.(From left to right, there are x-type intersection, y-type intersection, T-type intersection and irregularly shaped intersection.)}
\label{fig12}
\end{figure*}

\begin{figure*}[!t]
\centering
\includegraphics*[width=2.00in, height=2.25in, keepaspectratio=false]{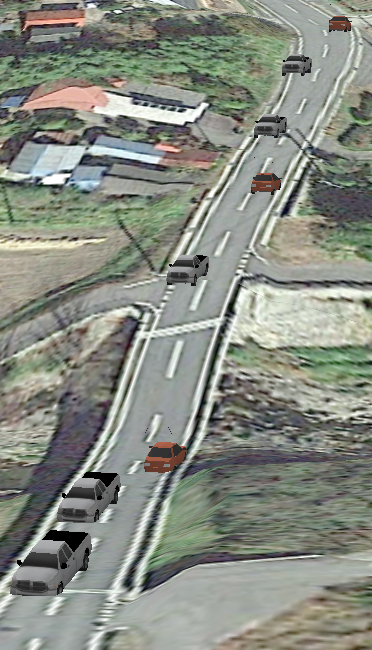}
\includegraphics*[width=2.85in, height=2.25in, keepaspectratio=false]{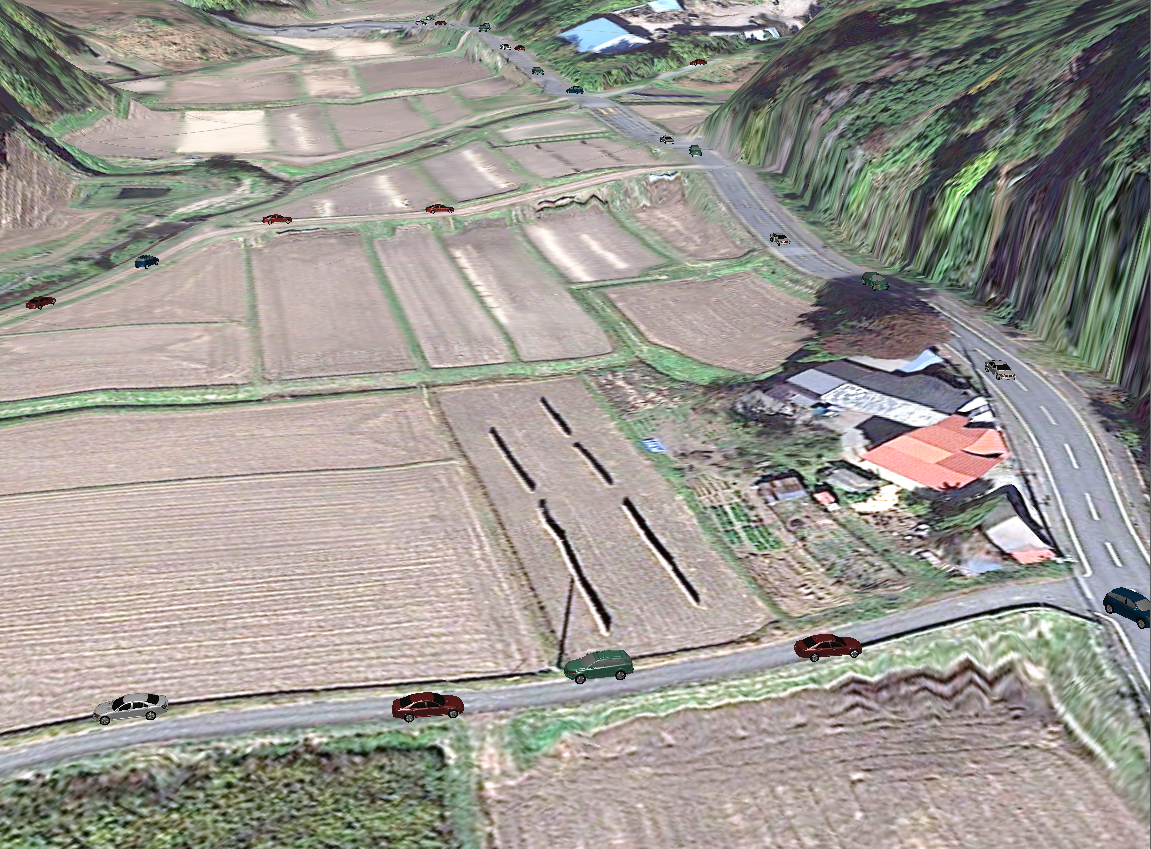}
\\
\vspace{+0.08cm}
\includegraphics*[width=4.9in, height=2.25in, keepaspectratio=false]{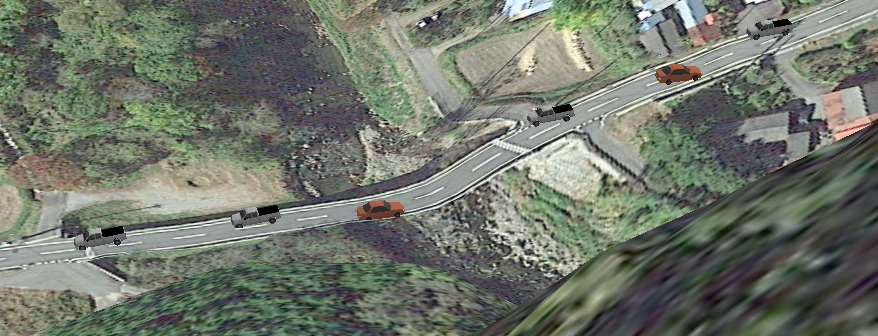}

\caption{Snapshots of traffic simulations based on the 3D road network}
\label{fig13}
\end{figure*}

Besides the Link and Intersection data, we also generated I\_Lane and L\_Lane for traffic simulations. Fig.\ref{fig12} shows the results of various intersections. The purple area is the Intersection geometry shape, the yellow lines are I\_Lane and L\_Lane. We can see that there is a high degree of consistency between I\_Lane and vehicle trajectories, and this consistency ensures the connectivity among road segments.
Based on the above scenario, we further simulate the road network in this scene. Fig.\ref{fig13} show some traffic simulation results based on the 3D roads.

\section{Discussion}
In our method, we use only GIS data as input to generate 3D network, and the roads in the generated scenes are high-detailed. They are flat and smooth, especially in the mountain area. In the urban area, the altitude difference is tiny (as shown in the last column of Fig.\ref{fig11}), therefore, our approach shows more advantages in countryside. Besides, we also support for generating various shapes of intersections, and our 3D road network can be used for traffic simulation. However, there are still some limitations of our method:
\begin{itemize}
\vspace{-0.08cm}
\item Due to the singularity of input data, we ignore the bridges, overpasses and tunnels, and they are important in urban traffic.
\vspace{-0.08cm}
\item The roundabout is not modelled when we generate the 2D road shape data. We will add the definition of roundabout in later scenes.
\vspace{-0.08cm}
\item When we generate 3D scenes, the edges of the roads are stretched because of the altitude difference between the road area and non road area. Using higher precision elevation data as input may solve this problem.
\end{itemize}

\section{Conclusion}
This paper proposes a method to generate large-scale 3D roads with road details by only using 2D road center axes, satellite imagery data and elevation data. By introducing a semantic description of the large-scale road networks and using image segmentations and a curves fitting method in the Frenet frame, our method generates the 3D shape of the road network and the traffic semantic structures, including the connectivity relationship, adjacency relationship and so on automatically. In the future, we would like to optimize the boundary treatment process in our 3D surface generation. Besides, we would also generate the traffic scenes according to the demand of crow simulation \cite{ACSEE2019,lv2018,chaochao2020,xu2018,xue2019,xumingliangSmog2018}.

\section*{Acknowledgments}
This work is supported and funded by the National Natural Science Foundation of China (grant no.62072415,62036010, and 61822701), the Natural Science Foundation of Henan Province (grant no.02300410496), the China Postdoctoral Science Foundation (grant no.2019MM662530), the foundation for the innovation of science and technology talents in higher education of Henan Province (grant no.18HASTIT020), and the special project for COVID-19 prevention and control emergency tackling of Henan Science and Technology Department (grant no. 201100312000). We would like to thank the reviewers for their helpful feedback.

\bibliographystyle{unsrt}

\bibliography{refs.bib}

\end{document}